# Conceptual Model with Built-in Process Mining

Sabah Al-Fedaghi
Computer Engineering Department
Kuwait University
Kuwait
sabah.alfedaghi@ku.edu.kw

*Abstract*—Process mining involves discovering, monitoring, and improving real processes by extracting knowledge from event logs in information systems. Process mining has become an important topic in recent years, as evidenced by a growing number of case studies and commercial tools. Current studies in this area assume that event records are created separately from a conceptual model (CM). Techniques are then used to discover missing processes and conformance with the CM, as well as for checks and enhancements. By contrast, in this paper we focus on modeling events as part of a tight multilevel CM that includes a static description, dynamics, events-log scheme, and monitoring and control system. If there is an out-of-model event log, it is treated as a requirement needed to build or enrich the CM. The motivation for such a unified system is our thesis that process mining is an essential component of a CM with built-in mining capabilities to perform self-process mining and attain completeness. Accordingly, our proposed conceptual model facilitates collecting data generated about itself. The resultant framework emphasizes an integrated representation of systems to include process-mining functionalities. Case studies that start with event logs are recast to evolve around a model-first approach that is not limited to the initial event log. The result presents a framework that achieves the aims of process mining in a more comprehensive way.

## I. Introduction

Process mining [1] is a branch of data science concerned with the handling of event records produced during the execution of organization processes. It involves discovering, monitoring, and improving real processes by extracting knowledge from event logs in information systems [2]. Process mining has become an important topic in recent years, as evidenced by a growing number of case studies and commercial tools, such as the site maintained by the IEEE Task Force on Process Mining [3][4].

Event logs that characterize behavior have been used in such areas as program visualization and concurrent-system analysis to infer an approximation model (see Fig. 1) that can be relied upon for creating a more complete CM. In this paper, *events* refers to "*activities* executed by resources at particular times and for a particular case" [5] (italics added). A model is a description that provides a reasonably rigorous specification (in this paper, a diagrammatic one) of the static structure and behavior of a system. The model is a depiction of what a system should be doing and what it is actually doing. Here, an explicit separation exists between description and execution. However, we mix the *models* used to enforce the *process execution* because they are necessarily synchronized. The execution is the activation of the model, and the model is a specification of the execution. We herein refer to processes occurring on a computer under the watchful eye of the system's monitoring component. Fig. 1 shows our vision of the place of the CM in a system.

Current process-mining studies assume that event records appear *separately* from model events (Fig. 2). The process-mining technique then tries to discover missing processes and conformance with the model, as well as for checks and enhancements. An independent log system (e.g., manual) collects the events data. By contrast, in the approach presented in this paper, we construct a *thinging machine* (TM) model by analyzing requirements, including possible non-model logs. The model automatically generates data about its events (see Fig. 3) as part of a tightly integrated model (see Fig. 4), model, then reexamining the model and its event logs is sufficient to make the model more complete. Such a procedure is similar to improving the dynamics of the model itself, such as changing the steps that are carried out in the model, and so on. This approach is presented as an alternative to a "wild-goose chase" effort to discover processes using an event-log system. Suppose that one stream of behavior is A→B→C. Trying to run B→C would be rejected because it is not an acceptable behavior (event stream). This is reported in the log component of the integrated model. Hence, the behavioral model may be modified to accept starting with B in addition to starting with A. Accordingly, the execution of the behavioral model would accept B→C as an acceptable stream of events. In this case, a missing process is discoverable through its rejection as reported in the log component of the system.

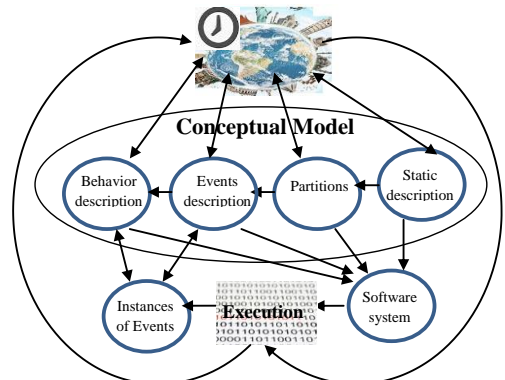

Fig. 1. General view of the conceptual model position between reality and software system.

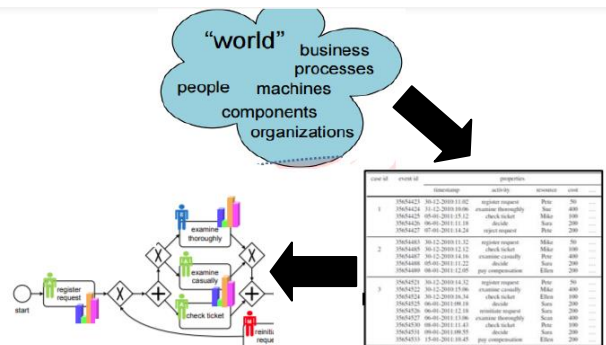

Fig. 2. Current visualization of process mining.

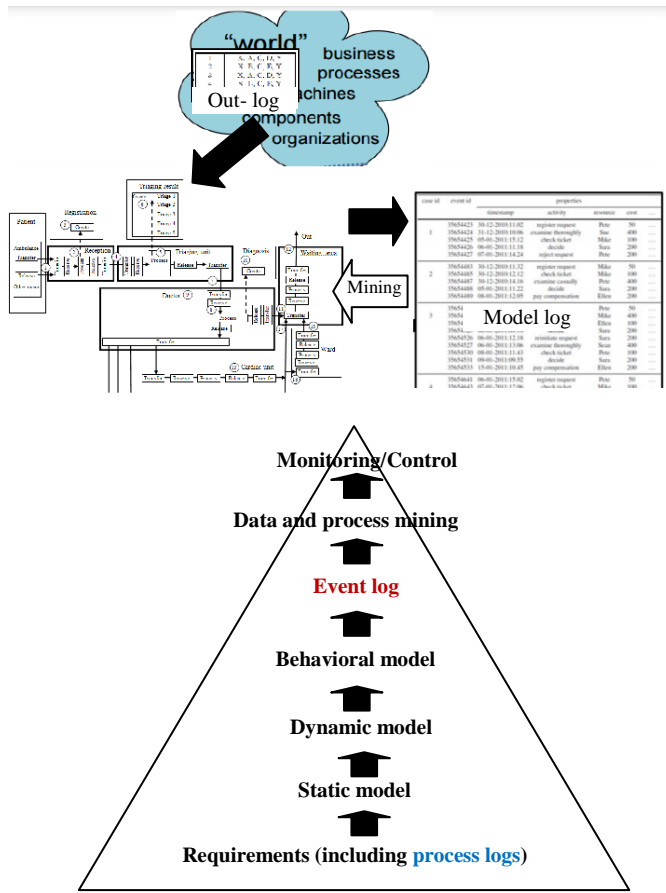

Fig. 4. The position of events in the system-development stages.

In section 2, we will briefly describe our main tool—that is, the TM model. The TM model has been applied in several diverse fields such as security [6] and privacy [7]. We provide a TM modeling example in Section 2 to clarify our notion of a conceptual model with built-in process mining. Section 3 applies our approach to a case study that is more complicated. Section 4 reviews related works.

The TM model involves a static model of the relationships between *things* (to be defined later) through *machine*s (to be defined later), a dynamic model of decompositions that embed behavior, event types, behavior in terms of chronology of events, an event log elicited from currently executed events, and a monitoring and control scheme that guides, enforces, or measures the execution. The motivation for such an integrated system is our thesis that if such an integrated model exists, it limits the need for model-less techniques for facilitating process-related problems (e.g., missing processes).

We claim that adopting an integral theoretical conceptual model takes care of tracing the process execution in the form of specifying all types of event streams (to be defined later). The events are generated by the event-log component as a part of the conceptual model function and not produced by an outside-log system. Note that the captured events in the log are already described in the *behavioral model* as some actions executed through time. The TM-based system can discover and treat issues such as a missing process.

The TM model includes only five generic actions that affect things: create, process, release, transfer, and receive. This specification contrasts with the ambiguous notion of activity (hence, the notion of event) used in current process-mining literature. If a process is missed in constructing such a

## II. TM Modeling

The TM model articulates the ontology of the world in terms of an entity that is simultaneously a *thing* and a *machine*, called a *thimac* [8-11]. A thimac is like a double-sided coin. One side of the coin exhibits the characterizations assumed by the thimac, whereas, on the other side, operational processes emerge that provide dynamics. A thing is subjected to doing, and a machine does. The simplest type of machine is shown in Fig. 5. The actions in the machine (also called stages) can be described as follows:

**Arrive:** A thing moves to a machine.
**Accept:** A thing enters the machine. For simplification, we assume that all arriving things are accepted; hence, we can combine the arrival and accept stages into one stage: the **receive** stage.
**Release:** A thing is ready for transfer outside the machine.
**Process:** A thing is changed, but no new thing results.
**Create:** A new thing is born in the machine.
**Transfer:** A thing is input into or output from a machine.

Additionally, the TM model includes storage and triggering (denoted by a dashed arrow in this study's figures), which initiates a flow from one machine to another. Multiple machines can interact with each other through movement of things or triggering. Triggering is a transformation from one series of movements to another.

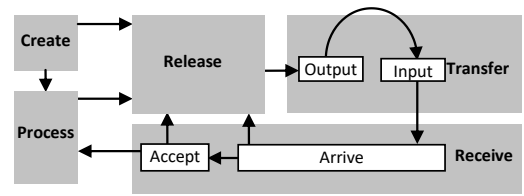

Fig. 5. The thinging machine.

**Example**: According to Weijters and Van der Aalst [12], "The models mined by process mining tools can be used as an objective starting point during the deployment of systems that support the execution of processes and/or as a feedback mechanism to check the prescribed process model against the enacted one." Weijters and Van der Aalst [12] illustrate how process-mining techniques work using an example of the event log shown in Fig. 6. This log shows the events involved in applying for a license to ride motorbikes or drive cars as follows:

X = Apply for license
A = Attend classes on how to ride motorbikes
B = Attend classes on how to drive cars
C = Do theoretical exam
D = Do practical exam to ride a motorbike
E = Do practical exam to drive a car
Y = Obtain result

Then, Weijters and Van der Aalst [12] construct a Petri net model that corresponds to the table in Fig. 7.

Instead, in TM, we consider the table in Fig. 6 to be collected data, along with other requirements gathered to develop the model of a license system. Thus, we minimally add new processes that make sense to achieve a reasonably complete model. Fig. 8 shows the resultant TM static model.

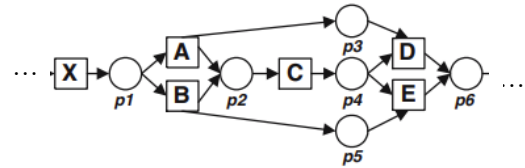

Fig. 6. Event log (adopted from [12]).

Fig. 7. Petri net model (partial adapted from [12]).

First, a person (circle 1) creates and sends an application to obtain a license (2). The application is received and is processed (3), and acknowledgement is sent to the applicant (4 and 5). The applicant (6) then attends classes on how to ride a motorbike (7) or how to drive a car (8).

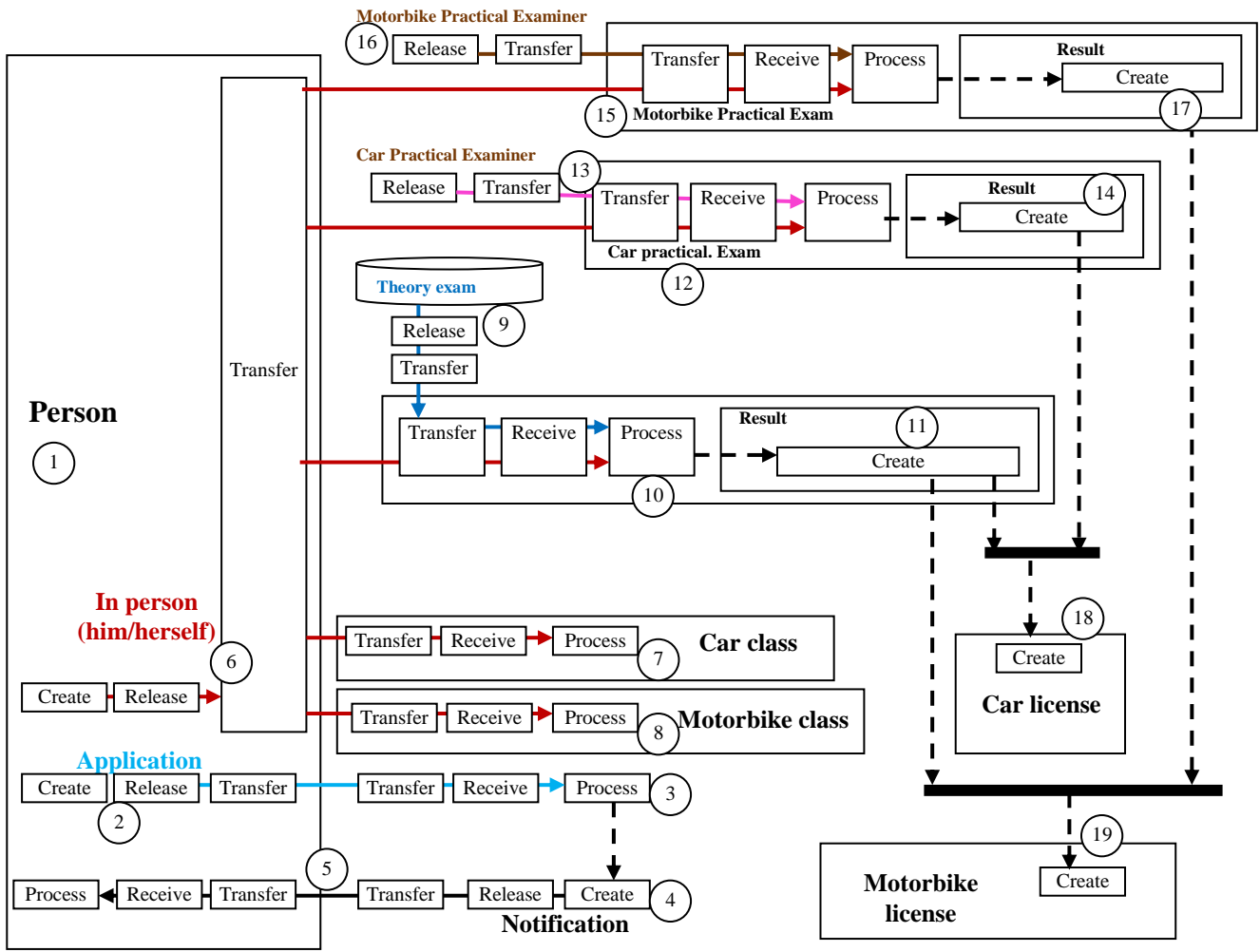

Fig. 8. The static TM model of the licensing system.

Next, the applicant takes the theoretical exam (9 and 10) that generates a result (11). Assuming the applicant passes the theoretical exam, he or she goes on to the practical driving exam (12) performed by an examiner (13), which produces a result (14). Alternatively, the applicant goes on to the practical riding exam (15) performed by an examiner (16), which produces a result (17). The results of the theoretical exam (11) and the practical driving exam (14) lead to a driver's license (18). The results of the theoretical exam (11) and the practical riding exam (17) lead to a motorcycle rider's license (19).

Such an approach is different from the process mining of Weijters and Van der Aalst [12] because it builds a complete model of the licensing system, which may use other typical requirement-collection methods. The next step in the TM approach is building the event-log scheme by finding all events in the static model of Fig. 8. This starts with identifying a set of events that are meaningful to the modeler. A TM event is defined based on (a) the region of an event, (b) the time of an event, and other attributes of events. Fig. 9 shows the representation of the event *A person applies for a license*. Accordingly, the static model is partitioned as shown in Fig. 10, where we assume that each partition (region) represents an event as follows.

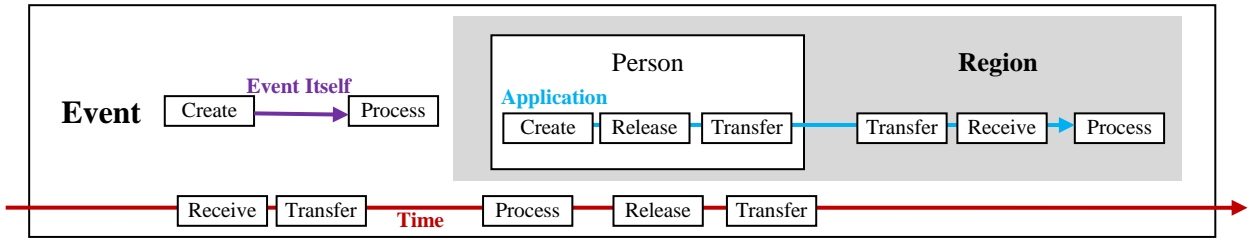

Fig. 9. The event *A person applies for a license*.

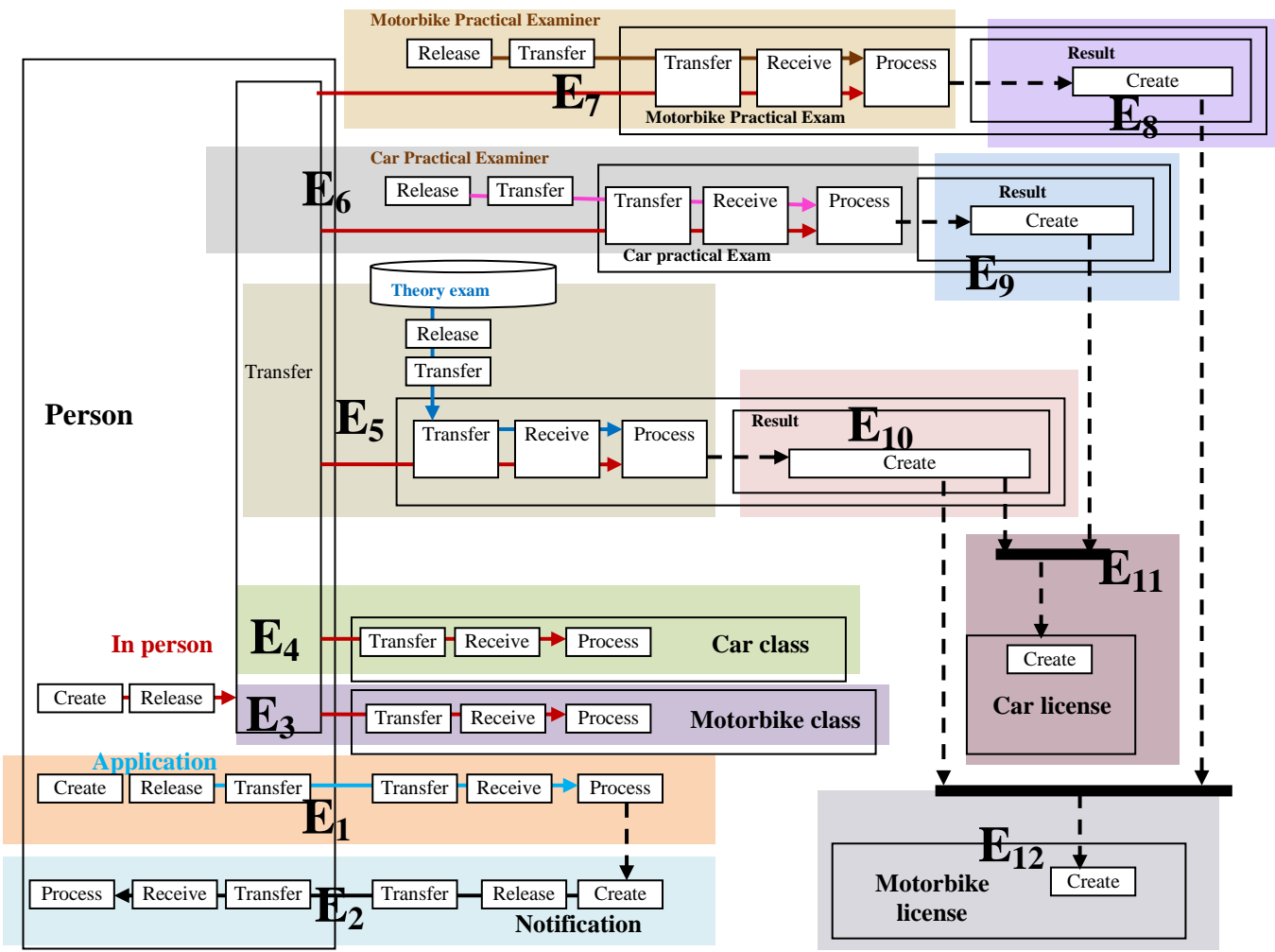

Fig. 10. The static TM model of the licensing system.

Event 1 ($E_1$): A person applies for a license.
Event 2 ($E_2$): An acknowledgement is sent to the applicant.
Event 3 ($E_3$): The applicant attends classes on how to ride motorbikes.
Event 4 ($E_4$): The applicant attends classes on how to drive a car.
Event 5 ($E_5$): The applicant takes the theoretical exam.
Event 6 ($E_6$): The applicant takes the practical driving exam.
Event 7 ($E_7$): The applicant takes the practical riding exam.
Event 8 ($E_8$): The result of the practical riding exam appears.
Event 9 ($E_9$): The result of the practical driving exam appears.
Event 10 ($E_{10}$): The result of the theoretical exam appears.
Event 11 ($E_{11}$): The applicant obtains a motorbike license.
Event 12 ($E_{12}$): The applicant obtains a car license.

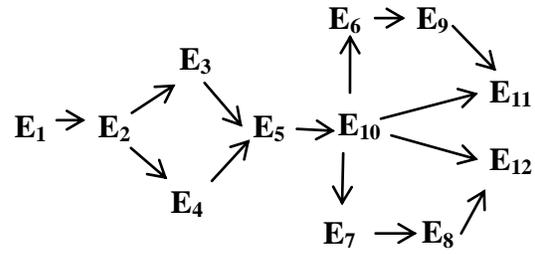

Fig. 11. The behavioral TM model of the licensing system.

Fig. 11 shows the behavioral model according to the chronology of events. At this stage, an event-log scheme can be developed to record each event. We call such a record a *meta event* as shown in Fig. 12. The set of meta events can be mined for various reasons, including process discovery. For example, suppose we have the event stream ($E_1$, $E_2$, $E_5$)—that is, an applicant applies for a license but then takes the theoretical exam without taking any classes. The monitoring system can easily recognize such a new behavior and reports it to the control system. Accordingly, a new process can be added to the behavioral model either automatically or manually (see Fig. 13). As another example, suppose that it is discovered from the event log that the theoretical exam and practical exam do not necessarily have to be in a particular order (e.g., a person can take the practical exam before the theoretical exam). Again, this can be discovered by mining the event log, and the behavioral model can be modified as shown in Fig. 14.

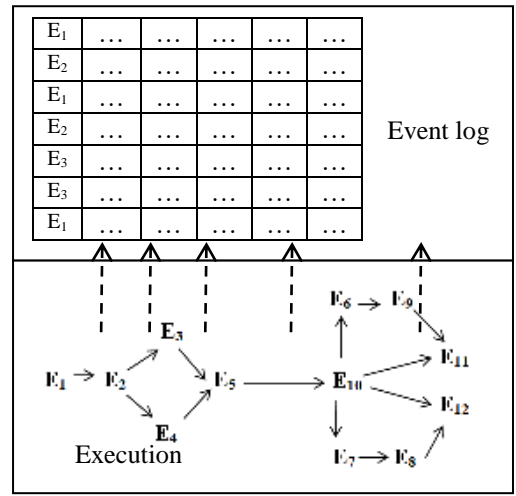

Fig. 12. The execution of the system generates meta events according to the behavioral model.

## III. HEALTH SYSTEM

In this section, we apply the TM approach introduced in the previous section to a large and real problem that involves health systems in four hospitals. According to Suriadi et al. [13] (see also Partington et al. [14]), variations in the treatment of patients across various hospitals substantially affect the quality and costs. The main research question is to identify the extent to which cross-hospital variations exist and why they exist. Suriadi et al. [13] used health care datasets to discover the pathways that patients traversed within hospitals. They compared process models and logs between various hospitals to identify subgroups (i.e., cluster of cases) that can explain the variations in patient flows.

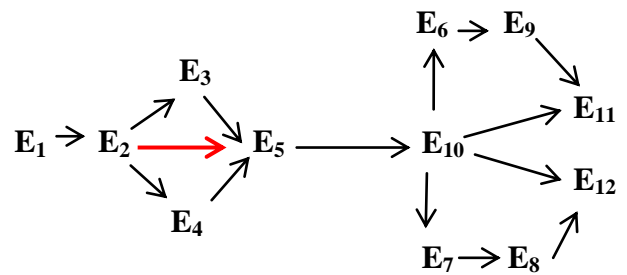

Fig. 13. The behavioral TM model of the licensing system such that a person can take the theoretical exam without taking

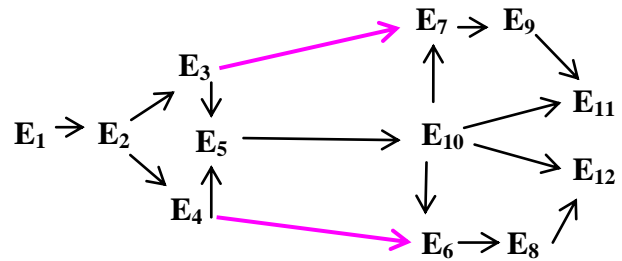

Fig. 14. The behavioral TM model of the licensing system where the theoretical exam and practical exams are not in any particular order.

In Suriadi et al.'s [13] case study, each hospital maintains an information system for managing operating theaters and tracking patient transfers between physical wards. The data extracted from these systems capture activities related to emergency department (ED) care. Suriadi et al. [13] excluded several cases (e.g., patient transfers and insufficiently documented cases).

A conceptual model is presented in terms of the UML class diagram (see partial view in Fig. 15). Despite the impressive work of Suriadi et al. [13] as a whole integration effort, we can see the typical assumption in many UML modeling projects in Fig. 15. Simply, the elementary conceptual notions are not in the right order. *Events*, an upper level notion, are mixed with static notions such as *patient* and *doctor*. As demonstrated in the previous section on the licensing system, time is a global feature that lifts the whole model from staticity to dynamicity. In the class diagram of Fig. 15, events are treated as a mere class.

In our recasting of this health system, the TM model includes all processes in every hospital as a holistic virtual description of the union of all ED processes. Each hospital schema reflects a partial view of this encompassing model. Thus, there are partial event logs in various hospitals. If there is a difference among various EDs, it is a subsystem variation (e.g., some hospitals do not provide some services in the ED). If an emergency process (say, p1) exists in Hospital A but not in Hospital B, then p1 can be discovered from comparing the (static) processes in the global conceptual model.

*A. Static Model*

As shown partially in Fig. 16, Partington et al. [14] used BPMN. Fig. 17 shows the holistic TM static model of the EDs as described in Partington et al. [14]. This model is supposed to be built upon inspection of each hospital's ED. Some details have been added to make the example more meaningful. In Fig. 17, a patient comes to the emergency unit by either an ambulance or other means (circle 1). In the reception, he or she is processed (2) to register the patient (3), and then he or she moves (4) to the triage unit where he or she is processed (5) to determine the degree of urgency (6). Accordingly, the patient then moves (7) to be processed (8) by a doctor (9) who writes a diagnosis (10). If some hospitals have additional processes (e.g., nurse processing), it is possible to add them to create a union for emergency operations that are not performed by Hospital 1.

Depending on the doctor's diagnosis, the patient moves to
- a waiting area (11) before leaving the hospital (12) or
- the cardiac, medical, A&E or other unit (13–16).

The patient either goes to the waiting area (17) before leaving the hospital or goes to a ward, and then he or she goes to the waiting area to leave (18–19).

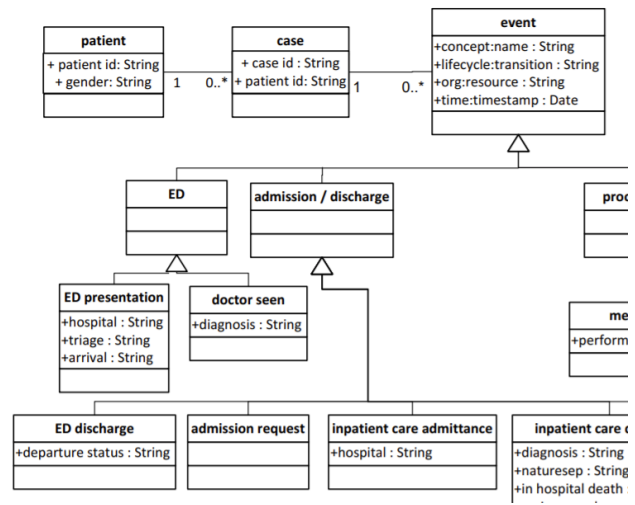

Fig. 15. Conceptual model of the event log used in Suriadi et al. [13] case study.

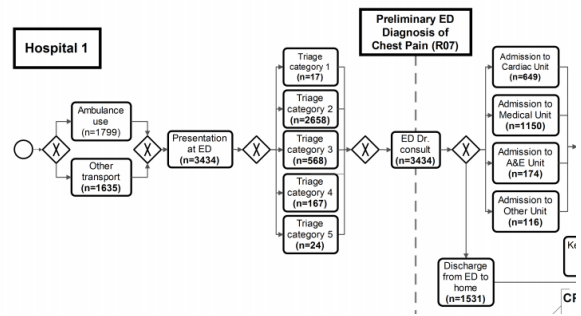

Fig. 16. Partial BPMN model of Hospital 1 (partial, from [14]).

*B. Dynamic Model*

At this stage, the modeling reaches a critical point that leads to defining what an event is. As mentioned in the licensing system in Section 2, an event in TM is a region in the static model that involves time and possibly other properties (not discussed in this ED description). Fig. 18 shows the event *The patient moves from the triage unit to be processed by a doctor*. Accordingly, the static model of Fig. 17 is divided into the decompositions shown in Fig. 19, where we represent each event by its region as follows:

Event 1 ($E_1$): A patient arrives at the ED by ambulance.
Event 2 ($E_2$): A patient arrives at the ED by other means.
Event 3 ($E_3$): The patient is received and is registered.
Event 4 ($E_4$): The patient moves to the triage unit.
Event 5 ($E_5$): The patient is processed in the triage unit.
Event 6 ($E_6$): The patient moves from the triage unit to a doctor.
Event 7 ($E_7$): A doctor examines the patient.
Event 8 ($E_8$): The patient leaves the doctor after being processed (i.e., diagnosed).
Event 9 ($E_9$): The patient goes to the waiting area and then leaves the ED.
Event 10 ($E_{10}$): The patient goes to the cardiac unit.

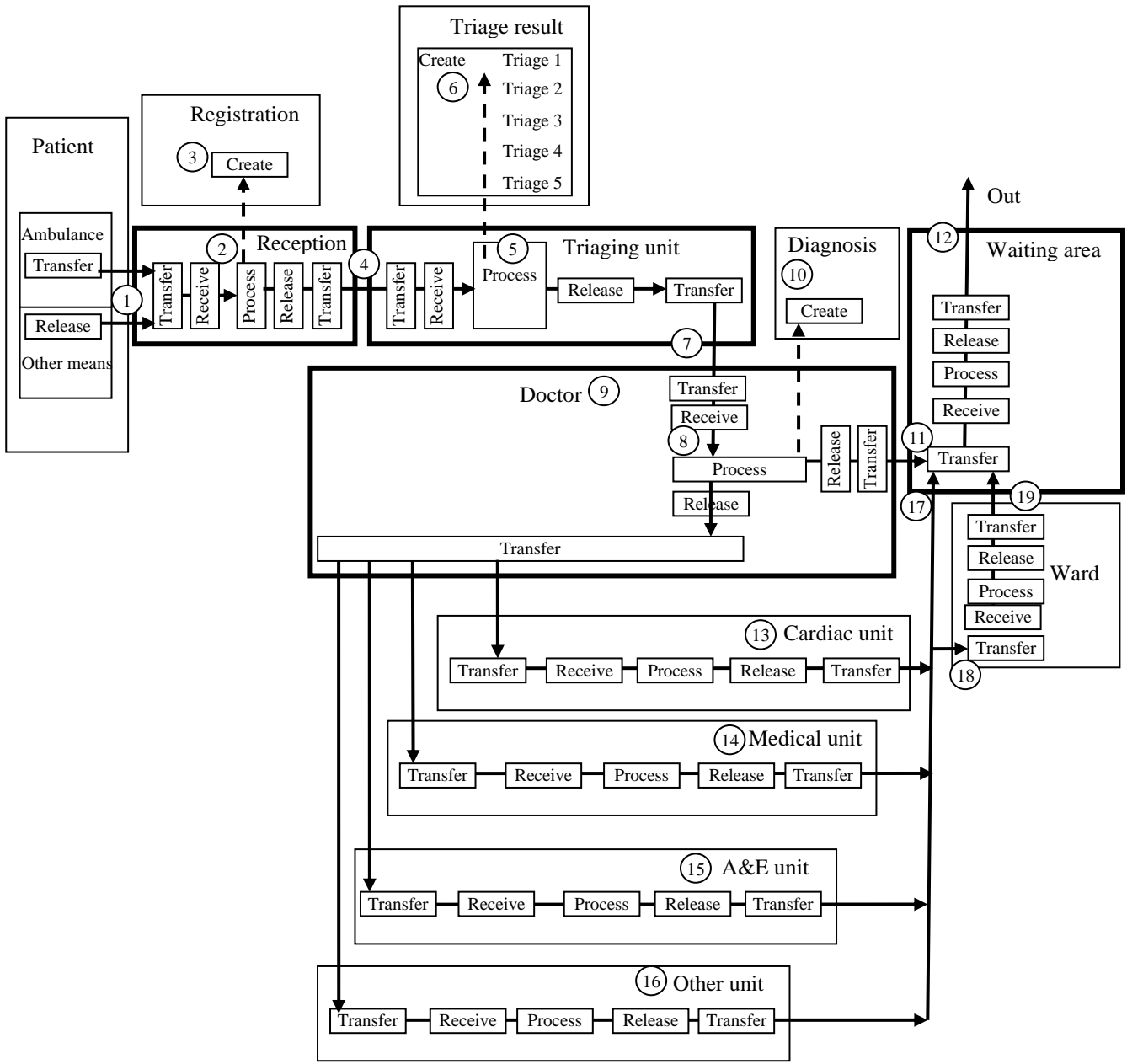

Fig. 17. The static model of the emergency department in a hospital.

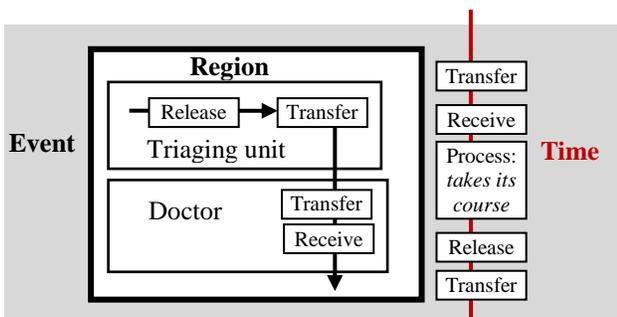

Fig. 18. The event *The patient moves from the triage unit to be processed by a doctor.*

Event 11 ($E_{11}$): The patient leaves the cardiac unit.
Event 12 ($E_{12}$): The patient goes to the medical unit.
Event 13 ($E_{13}$): The patient leaves the medical unit.
Event 14 ($E_{14}$): The patient goes to the A&E unit.
Event 15 ($E_{15}$): The patient leaves the A&E unit.
Event 16 ($E_{16}$): The patient goes to another unit.
Event 17 ($E_{17}$): The patient leaves the other unit.
Event 18 ($E_{18}$): The patient goes to the ward.
Event 19 ($E_{19}$): The patient dies in the ward.
Event 20 ($E_{20}$): The patient leaves the ward.

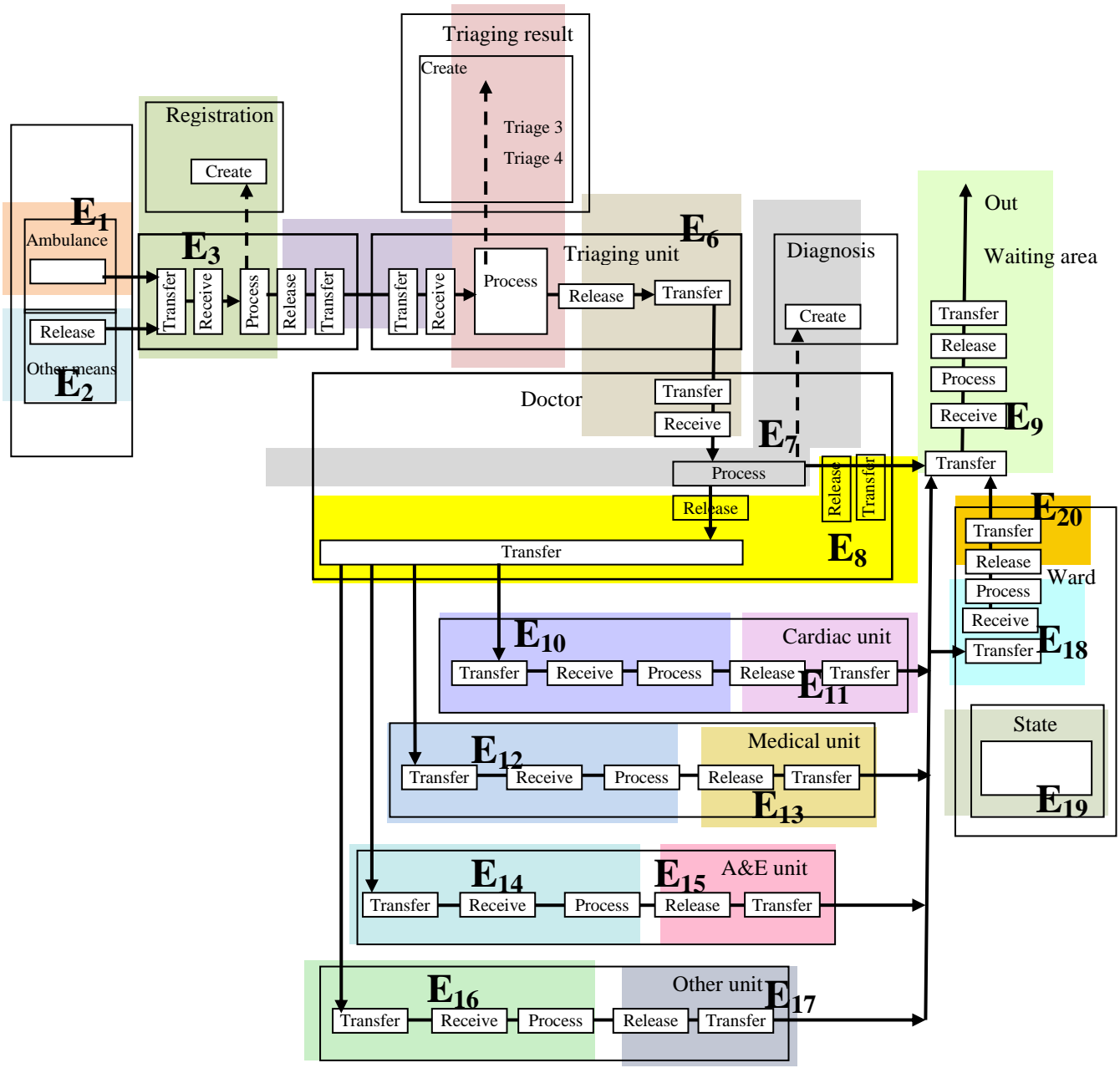

Fig. 19. The dynamic model of the emergency department in a hospital.

*C. Behavioral Model*

Fig. 20 shows the behavioral model in terms of the chronology of events. Each *stream* of events (the sequence of events for a single type of patient; e.g., E1, E3, E4, E5, E6, E7, E8, and E9) can be examined to see the process that a patient goes through. There are 40 types of event streams in Fig. 20. There are many *instances* of these types of streams. Any deviation from these streams results in alerts from the monitoring system. Note that each hospital has a sub-behavior of the global behavior. From such representation of events, we can discover a different or new ED behavior in one hospital, as shown in Fig. 21 (red arrows). In this case, the indicated hospital does not have an ambulance service. Additionally, a physician immediately examines the received patient; thus, an arrow that bypasses triaging is added. The point here is that with such a TM representation of the behavioral model, it is easier to discover missing processes. This development of a model is an alternative approach to chasing missing processes through the non-model-based event log. Thus, we expect that if all systems in the hospitals were remodeled using the TM model, the resultant behavioral representations would contrast with an overall model in the holistic system.

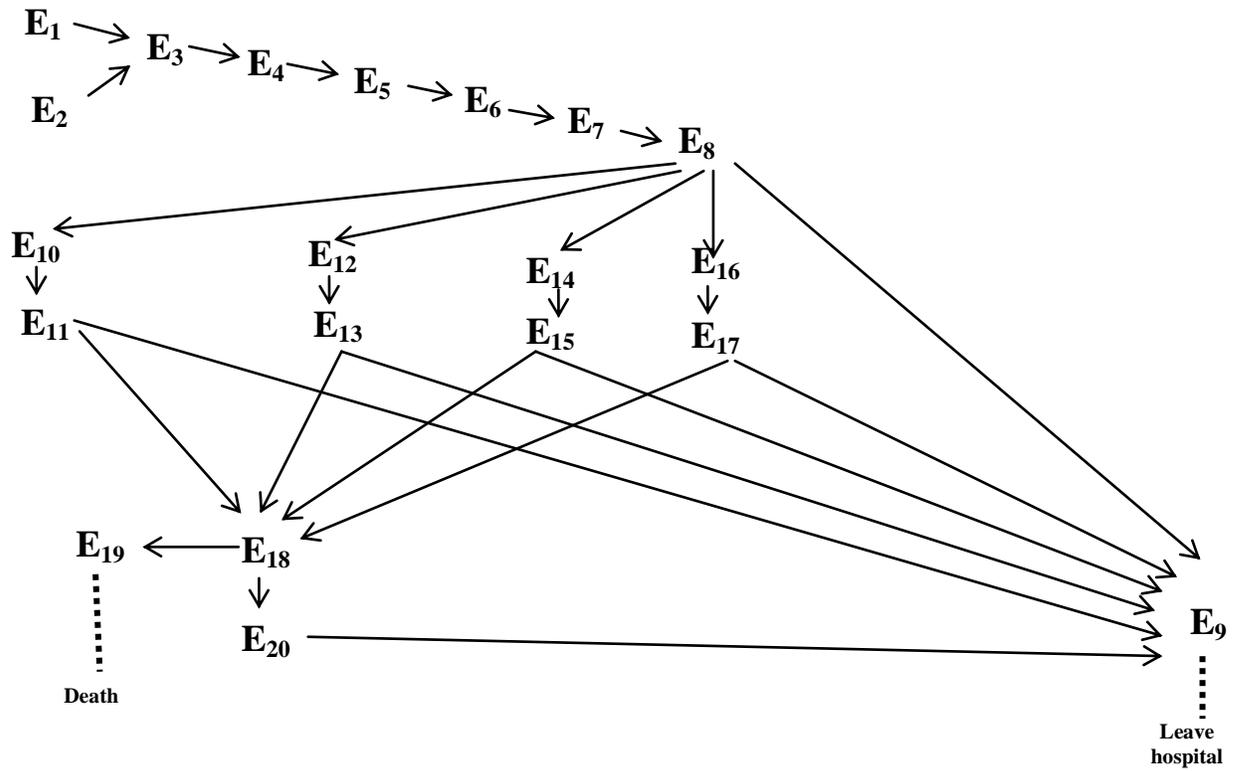

Fig. 20. The behavioral model of the emergency department in a hospital.

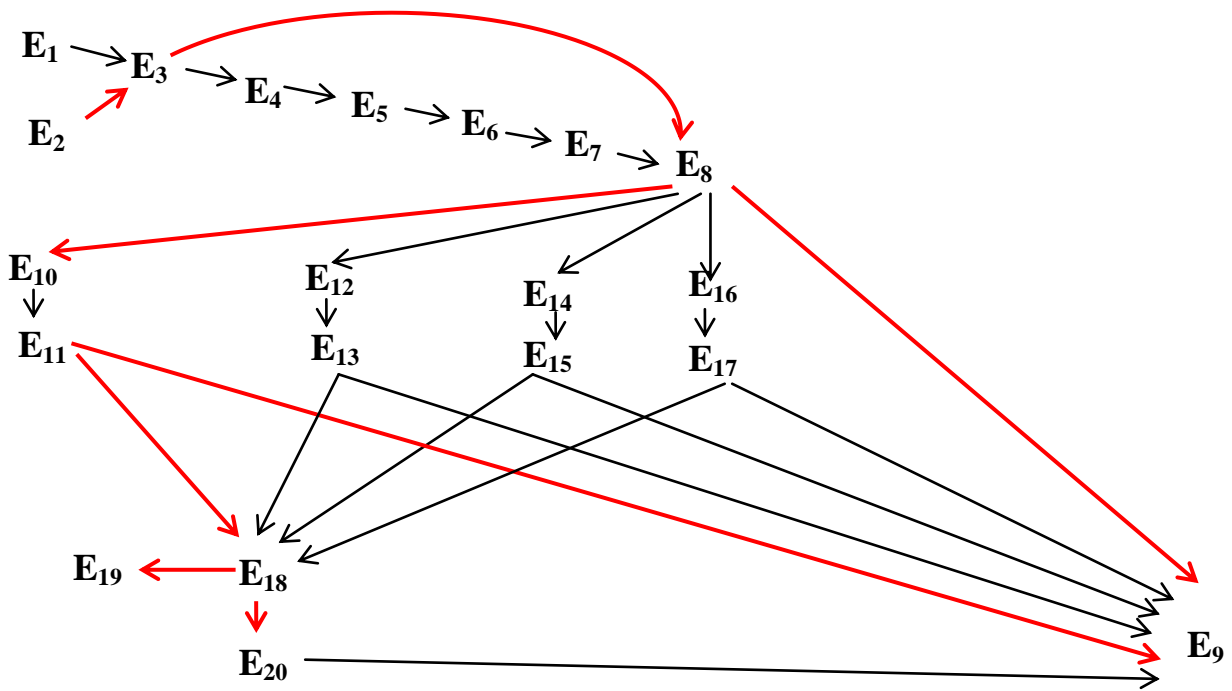

Fig. 21. Different behavioral model of the emergency department in a hospital.

## IV. Related Materials

Most of the information systems used by organizations do not record the execution data in a process-centric way so that the data are not ready for process mining. Techniques for event-log preparation can be categorized into methods for event data extraction, correlation, and abstraction. Techniques in this area assign semantics to data elements by defining how they can jointly be interpreted as the execution of a business process activity [15][16].

Process mining provides a set of techniques and algorithms for process discovery, conformance checking, and enhancement [1][17].

- Process discovery aims at the creation of a process model automatically from the data recorded during process execution [4].
- Conformance checking processes the recorded data based on a process model and provides diagnostic results [18].
- Process enhancement enriches a given process model based on the recorded data [19][20], thereby providing a more complete process representation.

According to van der Aalst [1], despite the maturity of the individual process-mining techniques, considerable resources have to be allocated in process-mining projects for the extraction and preparation of event data before the actual analysis can even start. Process-mining techniques use different representations and make different assumptions, and users often need to resort to trying different methods in an ad hoc manner [5]. Finding, merging, and cleaning event data remain a challenge for the application of process-mining techniques [2].

## V. Conclusion

In this paper, we have examined the notion of process mining. We proposed the conceptual TM model as a unifying description at the static, dynamic, and behavioral levels of the system with its own events-log component. Process mining takes place based on this event log of the system. A pre-model or out-of-model event log can be utilized in building the TM model. Once the model with its multilevel stages is built, then the model through its event-log component can mine its processes to discover new or missing processes that can be added, manually or automatically, to the specification of accepted behavior.

In this paper, we presented TM as a new model to be applied to process mining. Future research will elaborate on using TM in the process mining area with more complex examples. Specifically, TM needs to be related to such notions as process enhancement and conformance.